\documentclass[useAMS,usenatbib,usegraphicx]{mn2e}

\title[An analytical approximation of the growth function]
{An analytical approximation of the growth function in 
  Friedmann-Lema\^itre universes}
\author[M. Kasai]
{M. Kasai\thanks{E-mail: kasai@phys.hirosaki-u.ac.jp}\\
Graduate School of Science and Technology, Hirosaki University, 
Bunkyo-cho 3, Hirosaki 036-8561, Japan}
\begin{document}

\date{Accepted --- Received ---; in original form \today}

\pagerange{\pageref{firstpage}--\pageref{lastpage}} \pubyear{20??}

\maketitle

\label{firstpage}

\begin{abstract}
  We present an analytical approximation formula for the growth
  function in a spatially flat cosmology with dust and a cosmological
  constant.  Our approximate formula is written simply in terms of a
  rational function.  We also show the approximate formula in a dust
  cosmology without a cosmological constant, directly as a function of
  the scale factor in terms of a rational function.  The single
  rational function applies for all, open, closed and flat universes.
  Our results involve no elliptic functions, and have very small
  relative error of less than 0.2 per cent over the range of the scale
  factor $1/1000 \la a \lid 1$ and the density parameter $0.2 \la
  \Omega_{\rmn{m}} \lid 1$ for a flat cosmology, and less than $0.4$
  per cent over
  the range $0.2 \la \Omega_{\rmn{m}} \la 4$ for a
  cosmology without a cosmological constant.
\end{abstract}


\section{Introduction}

\cite{heath} has shown that the growth function in a dust cosmology can be
written as
\begin{equation}
  D_1(a) \propto H(a) \int_0^a \frac{da'}{\big(a' H(a')\big)^3},
\end{equation}
\begin{equation}
  D_2 \propto H(a), 
\end{equation}
\begin{equation}
  H(a) = \sqrt{\Omega_{\rmn{m}} a^{-3} +
    (1-\Omega_{\rmn{m}}-\Omega_{\Lambda}) a^{-2} 
    +\Omega_{\Lambda}}, 
\end{equation}
where $a$ is the scale factor of the universe, normalised to unity at
present,  $\Omega_{\rmn{m}}$ and
$\Omega_{\Lambda}$ is the density parameters of dust matter and a
cosmological constant, respectively. 

Although a compact expression using the incomplete beta function has
been shown by \cite{bbk}, so far, no analytic solution of $D_1(a)$ has
been presented for $\Omega_{\Lambda}\neq 0$.  Here we restrict
ourselves to the case $\Omega_{\rmn{m}}+\Omega_{\Lambda}=1$, and
present an approximate formula in a simple algebraic form.

Following \cite{eisen}, we adopt a normalization for $D_1(a)$ as
\begin{equation}
    D_1(a) =\frac{5\Omega_{\rmn{m}}}{2} H(a) \int_0^a \frac{da'}{\big(a'
      H(a')\big)^3}. 
\end{equation}
Then, in a flat cosmology $\Omega_{\rmn{m}}+\Omega_{\Lambda}=1$, our 
formula is
\begin{equation}\label{eq:4}
  D_1(a)=a\sqrt{1+x}\frac
  {1+1.175x+0.3064x^2+0.005355x^3}
  {1+1.857x+1.021x^2+0.1530x^3}, 
\end{equation}
where, 
\begin{equation}\label{eq:three}
  x = \frac{1-\Omega_{\rmn{m}}}{\Omega_{\rmn{m}}} a^3 . 
\end{equation}
Note that our approximate formula is  exact
when $\Omega_{\rmn{m}}=1$.

\section{Approximation}

We define 
\begin{equation}
  F(a) = \frac{5}{2} \Omega_{\rmn{m}}^{\frac{3}{2}} 
  \int_0^a \frac{da'}{\big(a' \sqrt{\Omega_{\rmn{m}} a'^{-3} +1-\Omega_{\rmn{m}}})\big)^3}. 
\end{equation}
The power series expansion of $F(a)$  around $a=0$
yields 
\begin{equation}
  F(a) = a^{\frac{5}{2}} \left(1 - \frac{15}{22} x
    + \frac{75}{136} x^2
    - \frac{175}{368} x^3
    + \cdots \right),  
\end{equation}
where $x$ is defined in Eq.~(\ref{eq:three}).  
After expanding $F(a)$ up to $O(x^6)$, we can obtain 
 the Pad\'e approximant to the following order:
 \begin{equation}
   F(a) = a^{\frac{5}{2}} \frac{1 + b_1 x + b_2 x^2 + b_3 x^3}
   {1 + c_1 x + c_2 x^2 + c_3 x^3}, 
 \end{equation}
where the numerical constants are determined as follows:
\begin{equation}
  b_1 = \frac{619226202351}{527102715964}, 
\end{equation}
\begin{equation}
  b_2 = \frac{12478731282519}{40730664415400}, 
\end{equation}
\begin{equation}
  b_3 = \frac{232758215919527}{43467765064114880}, 
\end{equation}
\begin{equation}
  c_1 = \frac{88964947071}{47918428724}, 
\end{equation}
\begin{equation}
  c_2 = \frac{2445658735707}{2395921436200}, 
\end{equation}
\begin{equation}
  c_3 = \frac{17010766061223}{111170754639680}. 
\end{equation}
We have checked the Taylor expansion and the Pad\'e approximant using
Maxima.\footnote{{\tt http://maxima.sourceforge.net/}}
It is convenient to use a normalised growth function 
\begin{equation}
  D(a) \equiv \frac{D_1(a)}{D_1(1)}, 
\end{equation}
so that D(1)=1. 

In order to check the accuracy of our formula, we calculate
the following relative error  
\begin{equation}
  \Delta E = \frac{|D^{\rmn{appr}} - D^{\rmn{num}}|}
  {D^{\rmn{num}}}\times 100 \ (\rmn{per cent}), 
\end{equation}
where $D^{\rmn{appr}}$ and $D^{\rmn{num}}$ represent the values of
normalised growth functions calculated by using approximate formula and
numerical method, respectively.

Table \ref{table:one} shows the maximal relative  error
$\Delta E$ for $1/1000 \lid a \lid 1$ 
in our method. It is apparent that our approximate formula has 
sufficiently small uncertainties over the wide range of parameters
$0.2 \lid \Omega_{\rmn{m}} \lid 1$ and $1/1000 \lid a \lid 1$. 
\begin{table}
  \caption{The maximal relative error $\Delta E$ (per cent) for the
    normalised growth function $D(a)$ by our formula Eq.~(\ref{eq:4}) for
    $\Omega_{\rmn{m}}+\Omega_{\Lambda}=1$}  
\label{table:one}
\begin{tabular}{cccccc}
\hline
$\Omega_{\rmn{m}}$ & 0.2 & 0.3 & 0.4 & 0.5 & 1 \\
\hline
maximal $\Delta E$ &  0.20 & 0.03 & $<0.01$ & $<0.01$ & 0 \\   
\hline 
\end{tabular}
\end{table}

The approximation of \cite{carroll}, which was adopted from
\cite{lahav},   is
\begin{equation}\label{eq:carroll}
  D_1^{\rmn{C}} = {\frac{5\Omega_{\rmn{m}}}{2}} \frac{1}{
    \Omega_{\rmn{m}}^{\frac{4}{7}} - (1-\Omega_{\rmn{m}}) +
    \left(1+\frac{\Omega_{\rmn{m}}}{2}\right)
    \left(1+\frac{1-\Omega_{\rmn{m}}}{70}\right) },  
\end{equation}
for $\Omega_{\rmn{m}} + \Omega_{\Lambda}=1$. 

A comparison of the relative error $\Delta E$ at $a=1$ for both
methods by \cite{carroll} and us, is shown in Table
\ref{table:carroll} for $\Omega_{\rmn{m}} + \Omega_{\Lambda}=1$. 
Our formula has generally smaller relative error in the range 
$0.2 \la \Omega_{\rmn{m}} < 1$. 
\begin{table}
  \caption{A comparison of the relative error 
 of Eq.~(\ref{eq:carroll}) and our $D_1(1)$ from Eq.~(\ref{eq:4}). }
\label{table:carroll}
\begin{tabular}{ccccc}
\hline
$\Omega_{\rmn{m}}$ & 0.2 & 0.3 &  0.5 & 0.9 \\
\hline
$\Delta E$ of Eq.~(\ref{eq:carroll}) 
&  0.54 & 0.134 &  0.057 & 0.019 \\   
$\Delta E$ of Eq.~(\ref{eq:4})
&  0.19 & $<0.01$ &  $<0.01$ & $<0.01$ \\   
\hline 
\end{tabular}
\end{table}

\section{Zero cosmological constant case}

In the $\Omega_{\Lambda}=0$ case, the analytical form for $D_1(a)$ is
widely studied (e.g., \citealt{wein}, \citealt{heath}). \cite{bbk} also
have shown the growth function as a function of the scale factor $a$
in various Friedmann-Lema\^itre universes, in a slightly different context.
Here, we show a simple and efficient approximate formula, directly as
a function of $a$.  In a dust cosmology with zero
cosmological constant, $\Omega_{\Lambda}=0$, our formula is
\begin{equation}\label{eq:17}
  D_1(a)=a\sqrt{1+y}\frac
  {1+1.113y +0.1247y^2-0.003893y^3}
  {1+2.185y+1.424y^2+0.2402y^3}, 
\end{equation}
where, 
\begin{equation}
  y = \frac{1-\Omega_{\rmn{m}}}{\Omega_{\rmn{m}}} a. 
\end{equation}
It is straightforward to derive Eq.~(\ref{eq:17}) from the Pad\'e
approximant, in the same way as shown in the previous section. 
Table~\ref{table:two} shows the maximal relative error of the our
formula for the normalised growth function $D(a)$ in the range $0.2
\lid \Omega_{\rmn{m}} \lid 4$ and $1/1000 \lid a \lid 1$.

\begin{table}
\caption{The maximal relative error $\Delta E$ for the
    normalised growth function $D(a)$ by our formula Eq.~(\ref{eq:17})
    for
    $\Omega_{\Lambda}=0$
}
\label{table:two}
\begin{tabular}{cccccc}
\hline
$\Omega_{\rmn{m}}$ & 0.2 & 0.5 & 1 &  2.0 &4.0 \\
\hline
maximal $\Delta E$ & 0.34 & 0.02 & 0 &   0.07& 0.36\\   
\hline 
\end{tabular}
\end{table}

\section{Conclusion}

We have presented a simple approximation formula for the
growth function in a spatially flat cosmology with dust and a
cosmological constant. Our formula is written in terms of a rational
function, and widely applicable over the range $1/1000 \la a \lid 1$
and $0.2 \la \Omega_{\rmn{m}} \lid 1$ with sufficiently small relative error
of less than $0.2$ per cent. 

We have also shown the approximate formula in a dust cosmology without
a cosmological constant in terms of a rational function. 
The single rational function applies for all, open, closed and flat
universes with relative error of less than $0.4$ per cent over the range 
$1/1000 \la a \lid 1$ and $0.2 \la \Omega_{\rmn{m}} \la 4$.

\label{lastpage}

\begin{thebibliography}{99}
\bibitem[\protect\citeauthoryear{Bildhauer et al.}{1992}]{bbk} 
Bildhauer S., Buchert T., Kasai M., 1992, A\&A, 263, 23
\bibitem[\protect\citeauthoryear{Carroll et al.}{1992}]{carroll} 
Carroll S. M., Press W. H., Turner E. L., 1992, ARA\&A, 30, 499
\bibitem[\protect\citeauthoryear{Eisenstein}{1997}]{eisen} 
Eisenstein D. J., 1997, preprint (astro-ph/9709054)
\bibitem[\protect\citeauthoryear{Groth \& Peebles}{1975}]{gp} 
Groth E. J., Peebles P. J. E., 1975, A\&A, 41, 143
\bibitem[\protect\citeauthoryear{Heath}{1977}]{heath} 
Heath D. J., 1977, MNRAS, 179, 351
\bibitem[\protect\citeauthoryear{Lahav et al.}{1991}]{lahav} 
Lahav O., Lilje P. B., Primack J. R., Rees M. J., 1991 MNRAS, 251, 129
\bibitem[\protect\citeauthoryear{Weinberg}{1972}]{wein} 
Weinberg S., 1972, Gravitation and Cosmology, Wiley, New York



\end{thebibliography}
\end{document}